\documentclass[10pt, letterpaper, twocolumn]{IEEEtran}
\usepackage[dvips,usenames]{color}
\usepackage{epsf}
\usepackage{times}
\usepackage{epsfig}
\usepackage{graphicx}
\usepackage{amsmath}
\usepackage{amssymb}
\usepackage{amsxtra}
\usepackage{here}
\usepackage{rawfonts}
\usepackage{times}
\usepackage{url}
\usepackage{cite}
\usepackage{multirow}
\usepackage[normalem]{ulem}
\usepackage{algorithmic}
\usepackage{algorithm}
\twocolumn
\newcommand{\diag}{{\mathrm{diag}}}
\newcommand{\rank}{{\mathrm{rank}}}

\newtheorem{definition}{\bf Definition}

\newlength{\aligntop}
\newlength{\alignbot}
\makeatletter

\begin{document}

\title{{Collaborative Spectrum Sensing from Sparse Observations in Cognitive Radio Networks}}

\author{\authorblockN{Jia (Jasmine) Meng$^1$, Wotao Yin$^2$, Husheng Li$^3$, Ekram Hossain$^4$, and Zhu Han$^1$\\}
\authorblockA{\small{
$^1$ Department of Electrical and Computer Engineering, University of Houston, USA\\}
$^2$ Department of Computational and Applied Mathematics, Rice University, USA\\
$^3$ Department of Electrical Engineering and Computer Science, University of Tennessee at Knoxville, USA\\
$^4$ Department of Electrical and Computer Engineering, University of Manitoba, Canada}
\thanks{A part of this work appeared in proceedings of IEEE ICASSP 2010.}
}
\date{}
\maketitle

%\begin{spacing}{2.0}
\begin{abstract}
Spectrum sensing, which aims at detecting spectrum holes, is the
precondition for the implementation of cognitive radio (CR).
Collaborative spectrum sensing among the cognitive radio nodes is
expected to improve the ability of checking complete spectrum
usage. Due to hardware limitations, each cognitive radio node
can only sense a relatively narrow band of radio spectrum. Consequently, the available
channel sensing information is far from being sufficient for
precisely recognizing the wide range of unoccupied channels. Aiming at breaking
this bottleneck, we propose to apply matrix completion and joint sparsity recovery to reduce
sensing and transmitting requirements and improve sensing results.
Specifically, equipped with a frequency selective filter, each
cognitive radio node senses linear combinations of
multiple channel information and reports them to the fusion center,
where occupied channels are then decoded from the reports by using
novel matrix completion and joint sparsity recovery algorithms. As a result, the
number of reports sent from the CRs to the fusion center is
significantly reduced.
We propose two decoding approaches, one based on matrix completion and the other based on joint sparsity recovery, both of which allow exact recovery from incomplete reports.
The  numerical results validate the effectiveness and robustness of our approaches. In particular, in small-scale networks, the matrix completion approach achieves exact channel detection with a number of samples no more than $50\%$ of the number of channels in the network,
while joint sparsity recovery achieves similar performance in large-scale networks.


\end{abstract}

%\newpage
%\setlength{\baselineskip}{25pt}
\section{Introduction}

Ever since the 1920s, every wireless system has been required to have an exclusive license from the government in order not to interfere with other users of the radio spectrum. Today, with the emergence of new technologies which enable new wireless services, virtually all usable radio frequencies are already licensed to commercial operators and government entities. According to former U.S. Federal Communications Commission (FCC) chair William Kennard, we are facing with a ``spectrum drought"~ \cite{SW04}. On the other hand, not every channel in every band is in use all the time; even for premium frequencies below 3 GHz in dense, revenue-rich urban areas, most bands are quiet most of the time. The FCC in the United States and the Ofcom in the United Kingdom, as well as regulatory bodies in other countries, have found that most of the precious, licensed radio frequency spectrum resources are inefficiently utilized~\cite{SH05,KS08}.

In order to increase the efficiency of spectrum utilization, diverse types of technologies have been deployed. Cognitive radio is one of those that leads to the greatest technological gain in wireless capacity. Through the detection and utilization of the spectra that are assigned to the licensed users but standing idle at certain times, cognitive radio acts as a key enabler for spectrum sharing. Spectrum sensing, aiming at detecting spectrum holes (i.e., channels not used by any primary users), is the precondition for the implementation of cognitive radio. The Cognitive Radio (CR) nodes must constantly sense the spectrum in order to detect the presence of the Primary Radio (PR) nodes and use the spectrum holes without causing harmful interference to the PRs. Hence, sensing the spectrum in a reliable manner is of vital importance and constitutes a major challenge in CR networks. However, detection is compromised when a user experiences shadowing or fading effects or fails in an unknown way. To get a better understanding of the problem, consider the following example: a typical Digital TV receiver operating in a 6 MHz band must be able to decode a signal level of at least -83 dBm without significant errors \cite{FCC04}. The typical thermal noise in such bands is -106 dBm. Hence a CR which is 30 dBm more sensitive has to detect a signal level of -113 dBm, which is below the noise floor~\cite{MSB06}. In such cases, one CR user cannot distinguish between an unused band and a deep fade. In order to combat such effects, recent studies suggest collaboration among multiple CR nodes for improving spectrum sensing performance.

Collaborative spectrum sensing (CSS) techniques are introduced to improve the performance of spectrum sensing. By allowing different secondary users to collaborate and share their information, PR detection probability can be greatly increased. CSS can be classified into two categories. The first category involves multiple users exchanging information \cite{GS05,LW08}, and the second category uses relay transmission \cite{GL07}. Some recent studies on collaborative spectrum sensing include cooperative scheme design guided by game theory \cite{GL05} and random matrix theory \cite{CDBN08}, cluster-based cooperative CSS \cite{ZZY05}, and distributed rule-regulated CSS \cite{CZ08}; studies concentrating on CSS performance improvement include \cite{ZL08} introducing spatial diversity techniques to combat the error probability due to fading on the reporting channel between the CR nodes and the central fusion center. There are also studies concerning other interesting aspects of CSS performance under different constraints \cite{CDBN08,SZL07,LZPH08,ZMLKS10}. Very recently, there are emerging applications of the compressive sensing concept for CSS \cite{Tian08}.

Existing literature mostly focuses on the CSS performance examination when the centralized fusion center receives and combines {\em all} CR reports. In an $n$ channel cognitive radio network with $m$ CR nodes, the fusion center has to deal with $n*m$ reports and combine them wisely to form a channel sensing result. However,it is known that wireless channels are subject to fading and shadowing. When secondary users experience multi-path fading or happen to be shadowed, the reports transmitted by CR users are subject to transmission loss. As a result, in practice, no entire report data set is available at the fusion center. Besides, due to the fact that each cognitive radio can only sense a small proportion of the spectrum with limited hardware, each CR user gathers only very limited information about the entire spectrum.

\textbf{Contributions:}

We seek to release CRs from sending, and the central control unit from gathering, an excessively large number of reports, also target at the situations where there are only a few CR nodes in a large network and thus unable to gather enough sensing information for the traditional CSS. We propose to equip each cognitive radio node with a frequency selective filter, which linearly combines multiple channel information. The linear combinations are sent as reports to the fusion center, where the occupied channels are  decoded from the reports by compressive sensing algorithms. As a consequence, the amount of channel sensing at CRs and the number of reports sent from the CRs to the fusion center are both significantly reduced.

Following our previous work \cite{Jia1,Jia2} on compressive sensing, we propose two approaches to collaborative spectral sensing. The first approach is based on solving a matrix completion problem \cite{CR08,CCS08,MGC08,KO09,DM09}, which seeks to efficiently reconstruct a matrix (typically  low-rank) from a relatively small number of revealed entries. In this approach, the entries of the underlying matrix are linear combinations of channel powers. Each CR node takes its local spectrum measurements, but instead of directly recording channel powers, it uses its  frequency-selective filters to take $p$ \emph{linear combinations} of channel powers and reports them to the fusion center. The total $p\times m$ linear combinations taken by $m$ CRs form a $p\times m$ matrix at the fusion center. Considering transmission loss, we allow the the matrix to be incomplete. We show that this matrix is low-rank and has the properties enabling its reconstruction from only a small number of its entries, and therefore, information about the complete spectrum usage can be recovered from a small number of reports from the CR nodes. This approach significantly reduces the amount of sensing and communication workload.

The second approach is based on joint sparsity recovery \cite{DSWBB,TGS05,FR08,NT08,LD05}, which is motivated by the observation that the spectrum usage information the CR nodes collect has a common sparsity pattern: each of the few occupied channels is typically observed by multiple CRs. We develop a novel algorithm for joint sparsity signal recovery, which is more effective than existing algorithms in the compressive sensing literature since it can accommodate a large dynamic range of channel gains.

In both approaches, every CR senses all channels (by taking random linear projections of the powers of all channels), and the CRs do not communicate. While they work independently, their measurements are analyzed jointly by the detection algorithms running at the fusion center. Therefore, our approaches are very different from the existing collaborative spectrum sensing schemes in which different CRs are assigned to different channels. Our approaches move from collaborative sensing to ``collaborative" computation and shift coordination from the sensing phase to the post-sensing phase.

Our work is among the first that applies matrix completion or joint sparsity recovery to collaborative spectrum sensing in cognitive radio networks. Matrix completion and joint sparsity recovery are both being intensively studied in the compressive sensing community. We present them both because it is too early at this time to make a verdict of an eventual winner.

The rest of this paper is organized as follows: In Section \ref{sec:model}, the system model is given.
The matrix completion-based algorithm for collaborative sensing is described in Section \ref{sec:algorithm1}, and the joint sparsity based algorithm is described in Section \ref{sec:algorithm2}. After that, in Section \ref{sec:Discussion} we compare the two proposed approaches, discuss their computational complexity as well as filter design and dynamic update. Simulation results are presented in Section \ref{sec:simulation}, and conclusions are drawn in Section \ref{sec:conclusion}.

\section{System Model}\label{sec:model}

We consider a cognitive radio network with $m$ CR nodes that locally monitor a subset of $n$ channels. A channel is either occupied by a PR or unoccupied, corresponding to the states $1$ and $0$, respectively. We assume that the number $s$ of occupied channels is much smaller than $n$. The goal is to recover the occupied channels from the CR nodes' observations. Since each CR node can only sense limited spectrum at a time, it is impossible for limited $m$ CRs to observe  $n$ channels  simultaneously.

\begin{center}
\begin{figure}
  \center
  \includegraphics[width=3.7in,height=2.7in]{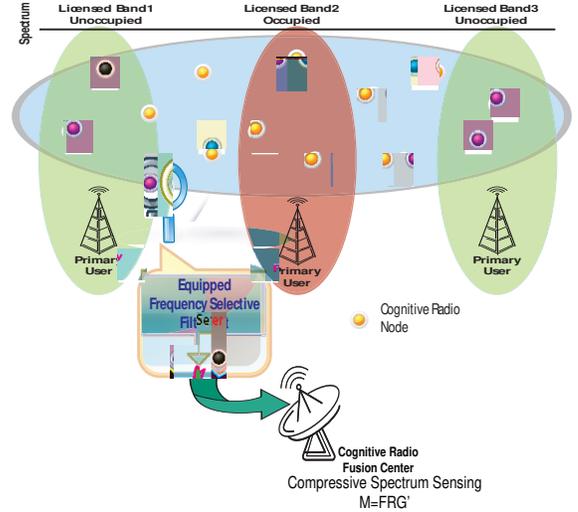}\\
  \caption{System model.}\label{f:system_model}
\end{figure}
\end{center}

To overcome this problem, we propose the scheme depicted in Fig. \ref{f:system_model}. Instead of scanning all channels and sending each channel's status to the fusion center,  using its
frequency-selective filters, a CR takes a small number of measurements that are linear combinations of multiple channels. The filter coefficients can be designed and implemented easily. In order to mix the different channel sensing information, the filter coefficients are designed to be random numbers. Then, these filter outputs are sent to the fusion center. Suppose that there are $p$ frequency selective filters in each CR node sending out $p$ reports regarding the $n$ channels. For the non-ideal cases, where we have relatively less measurements $pm<n$, i.e., the number of reports sent from all CRs is less than the total number of channels. The sensing process at each CR can be represented by a $p\times n$ filter coefficient matrix $\mathbf{F}$. Let an $n \times n$ \emph{diagonal} matrix $\mathbf{R}$ represent the states of all the channel sources using $0$ and $1$ as diagonal entries, indicating the unoccupied or occupied states, respectively. There are $s$ nonzero entries in $\diag(\mathbf{R})$. In addition, channel gains between the CRs and channels are described in an $m \times n$ channel gain matrix $\mathbf{G}$ given by \cite{RBook2002}:
\begin{equation}\label{eqn:G}
\mathbf{G}_{i,j}=P_i(d_{i,j})^{-\alpha/2}|h_{i,j}|
\end{equation}
where $P_i$ is the $i^{th}$ primary user's transmitted power, $d_{i,j}$ is the distance between the primary transmitter using $j^{th}$ channel and the $i^{th}$ CR node, $\alpha$ is the propagation loss factor, and $h_{i,j}$ is the channel fading gain. For AWGN channel, $h_{i,j}=1, \forall i,j$; for Rayleigh channel, $|h_{i,j}|$ follows independent Rayleigh distribution; and for shadowing fading, $|h_{i,j}|$ follows log-normal distribution \cite{RBook2002}. Without loss of generality, we assume that all PRs' use unit transmit power (otherwise, we can compensate by altering the corresponding channel gains). The measurement reports sent to the fusion center can be written as a $p\times m$ matrix
\begin{equation}
\mathbf{M}_{p \times m}=\mathbf{F}_{p \times n}\mathbf{R}_{n \times n}(\mathbf{G}_{m \times n})^\top.
\end{equation}
Note that due to loss or errors, some of the entries of $\mathbf{M}$ are possibly missing. The binary numbers on the diagonal of $\mathbf{R}$ are the $n$--channel states that we shall estimate from the available entries of $\mathbf{M}$.

\section{CSS Matrix Completion Algorithm}\label{sec:algorithm1}
It is typically difficult for the fusion center to acquire all entries of $\mathbf{M}$ due to transmission failure, which means that our observation is a subset $\mathbf{E}\subseteq [p]\times[m]$ of $\mathbf{M}$. However, it is possible to recover the missing entries in $\mathbf{M}$ since it holds the following two important properties \cite{CR08} required for matrix completion:
\begin{enumerate}
\item \textbf{Low Rank}: $\rank(M)$ equals to $s$, which is the number of prime users in the network and is usually very small.
\item \textbf{Incoherent Property}: Generate $\mathbf{F}$ randomly (subject to hardware limitation). From \eqref{eqn:G} and the fact that $\mathbf{R}$ has only $s$ nonzeros on the diagonal, $\mathbf{M}$'s SVD factors $\mathbf{U}$, $\Sigma$, and $\mathbf{V}$ satisfy the {\em incoherence condition} \cite{KO09}.
\begin{itemize}
  \item There exists a constant $\mu_{0}> 0$ such that for all $i \in [p]$, $j \in [m]$, we have $\sum_{k=1}^{s} \mathbf{U}_{i,k}^{2} \leq \mu_{0} s$, $\sum_{k=1}^{s} \mathbf{V}_{i,k}^{2}$ $\leq \mu_{0} s$.
  \item There exists $\mu_{1}$ such that $\mid \sum_{k=1}^{s} \mathbf{U}_{i,k}\Sigma_{k} \mathbf{V}_{j,k} \mid \leq \mu_{1} s^{1/2}$.
\end{itemize}
\end{enumerate}

$\mathbf{M}$ is in general incomplete because of transmission failure. Moreover, each CR might only be able to collect a random (up to $p$) number of reports due to the hardware limitation.
Therefore, the fusion certain receives a subset $\mathbf{E}\subseteq [p]\times[m]$ of $\mathbf{M}$'s entries.
We assume that the received entries are uniformly distributed with high probability\footnote{Depending on the different channel gain, the CRs will select different coding/modulation/power control schemes so that the received signal to noise ratio can be maintained about a certain threshold. Due to this reason, we can assume that the loss of information is uniformly distributed.}. Hence, we work with a model in which each entry shows up in $\mathbf{E}$ identically and independently with probability $\epsilon /\sqrt{p\times m}$.
Given $\mathbf{E}_{p\times m}$, the partial observation of $\mathbf{M}$ is defined as a $p\times m$ matrix given by
\begin{equation}
{M}_{ij}^{E}=\left\{
\begin{array}{ll}
{M}_{ij}, & \mbox{if} \ (i,j)\in \mathbf{E}\\
0, & \mbox{otherwise}.
\end{array}
\right.
\end{equation}
We shall first recover the unobserved elements of $\mathbf{M}$ from $\mathbf M^E$. Then, we reconstruct $(\mathbf{R}\mathbf{G}^\top)$ from the given $\mathbf{F}$ and $\mathbf{M}$ using the fact that all but $s$ rows of $(\mathbf{R}\mathbf{G}^\top)$ are zero. These nonzero rows correspond to the occupied channels.
Since $p$ and $m$ are much smaller than $n$, our approach requires a much less amount of sensing and transmission, compared to traditional spectrum sensing in which each channel is monitored separatively.

In previous research on matrix completion \cite{CCS08, MGC08, KO09, DM09}, it was proved that under some suitable conditions, a low-rank matrix can be recovered from a random, yet small subset of its entries by nuclear norm minimization:
\begin{equation}\label{mtx_cmp}
\min_{\mathbf{M}\in\mathbb{R}^{p\times n}} \tau\|\mathbf{M}\|_{\ast}+\frac{1}{2} \sum_{(i,j)\in\mathbf{E}}\left|\mathbf{M}_{i,j}-\mathbf{M}^{E}_{i,j}\right|^2
\end{equation}
where $\|\mathbf{M}\|_\ast$ denotes the nuclear norm of matrix $\mathbf{M}$ and $\tau$ is a parameter discussed in Section \ref{sec:stop} below. For notational simplicity, we introduce the linear operator $\mathcal{P}$ that selects the components $\mathbf{E}$ out of a $p\times n$ matrix and form them into a vector such that $\|\mathcal{P}\mathbf{M}-\mathcal{P}\mathbf{M}^{E}\|^2_2=\sum_{(i,j)\in\mathbf{E}}| \mathbf{M}_{i,j}-\mathbf{M}^{E}_{i,j}|^2$. The adjoint of $\mathcal{P}$ is denoted by $\mathcal{P}^*$.

Recent algorithms developed for (\ref{mtx_cmp}) include, but not limited to, the singular value thresholding (SVT) algorithm
\cite{CCS08} and the fixed-point continuation iterative algorithm
(FPCA) \cite{MGC08} for fast completion of large-scale matrices
(e.g., more than $1000$$\times$$1000$), a special trimming step
introduced by Keshavan et al. in \cite{KO09}. % The motivation is that although sampling is carried out uniformly at random, there is possibility that the matrix $\mathbf{M}^{E}$ contains columns and rows with $\Theta(\log(m)/\log\ \log(m))$ revealed entries. The largest singular values of $\mathbf{M}^{E}$ are of the order of $\Theta(\sqrt{\log(m)/\log\ \log(m)})$. The corresponding singular vectors are highly concentrated on high-weight column or row indices, respectively for left and right singular vectors. Such singular vectors offer artificially high weights columns/rows and thus are not able to provide useful information about the hidden entries of $\mathbf{M}$. By trimming, we set to zero all columns in $\mathbf{M}^{E}$ with number of none-zero entries greater than $2\mid \mathbf{E}\mid/n$. Set to zero all rows in $\textbf{M}^{E}$ with a number of none-zero entries greater than $2\mid \textbf{E}\mid/m$. Trimming helps getting rid of the artificially high weights columns/rows, and makes a clearer reveal of the low rank matrix structure.

For our problem, we adopt FPCA, which appears to run very well for our small--dimensional tests.
In the following subsections, we describe this algorithm and the steps we take for nuclear norm minimization. Also, we study how to use the approximate singular value decomposition (SVD)-based iterative algorithm introduced in \cite{MGC08} for fast execution. We further discuss the stopping criteria for iterations to acquire optimal recovery. Finally we show how to obtain $\mathbf{R}$ from the estimation $\tilde{\mathbf{M}}$ of $\mathbf{M}$.

\subsection{Nuclear Norm Min. via Fixed Point Iterative Algorithm}

FPCA is based on the following fixed--point iteration:
\begin{equation}\label{eqn:steps}
\left\{
\begin{array}{ll}
\mathbf{Y}^{k}=\mathbf{M}^{k}-\delta_{k}\mathcal{P}^*(\mathcal{P}\mathbf{M}^k-\mathcal{P}\mathbf{M}^{E}) \\
\mathbf{M}^{k+1}=S_{\tau\delta_{k}}(\mathbf{Y}^{k})
\end{array}
\right.
\end{equation}
where $\delta_{k}$ is step size and $S_\alpha(\cdot)$ is the matrix shrinkage operator defined as follows:
\begin{definition}\label{def:shrinkage}
\textbf{Matrix Shrinkage Operator $S_{\alpha}(\cdot)$}:
Assume $\mathbf{M} \in \mathbb{R}^{p\times m}$ and its SVD is given by $\mathbf{M}=\mathbf{U}\diag(\sigma)\mathbf{V}^T$, where $\mathbf{U} \in \mathbb{R}^{p\times r}$, $\sigma \in\mathbb{R}_{+}^{r}$, and $\mathbf{V} \in \mathbb{R}^{m\times r}$. Given $\alpha>0$, $S_{\alpha}(\cdot)$ is defined as \begin{equation}
S_{\tau}(\mathbf{M}):= \mathbf{U}\diag\left(s_{\alpha}(\sigma)\right)\mathbf{V}^{T}
\end{equation}
with the vector $s_{\alpha}(\sigma)$ defined as:
\begin{equation}
s_{\alpha}(x):=\max\{x-\alpha,0\},~\mbox{component-wise.}
\end{equation}
\end{definition}
Simply speaking, $S_{\tau}(\mathbf{M})$ reduces every singular values (which is nonnegative) of $\mathbf{M}$ by $\tau$; if one is smaller than $\alpha$, it is reduced to zero. In addition, $S_{\alpha}(\mathbf{M})$ is the solution of
\begin{equation}\label{eq:optimization_eq}
\min_{\mathbf{X}\in\mathbb{R}^{m\times n}} \alpha\|\mathbf{X}\|_{\ast}+\frac{1}{2} \|\mathbf{X}-\mathbf{M}\|_{F}^{2}
\end{equation}
where $\|\cdot\|_F$ is the Frobenius norm.

To understand \eqref{eqn:steps}, observe that the first step of \eqref{eqn:steps}  is a gradient-descent applied to the second term in \eqref{mtx_cmp} and thus reduces its value. Because the previous gradient-descent generally increases the nuclear norm, the second step of \eqref{eqn:steps} involves solving \eqref{eq:optimization_eq} to reduce the nuclear norm of $\mathbf{Y}^{k}$. Iterations based on \eqref{eqn:steps} converge when the step sizes $\delta_{k}$ are properly chosen (e.g., less than 2, or select by line search) so that the first step of \eqref{eqn:steps} is not ``expansive'' (the other step is always non-expansive).

\subsection{Approximate SVD Based Fixed Point Iterative Algorithm}

As stated in \cite{MGC08}, the second step of \eqref{eqn:steps} requires computing the SVD decomposition of $\mathbf{Y}^{k}$, which is the main computational cost of \eqref{eqn:steps}. However, if one can predetermine the rank of the matrix $\mathbf{M}$, or have the knowledge of the approximate range of its rank, a full SVD can be simplified to computing only a rank-$r$ approximation to $\mathbf{Y}^{k}$. Combined with the above fixed point iteration, the resulting algorithm is called fixed-point continuation algorithm with approximate SVD (FPCA). Specifically, the approximate SVD is computed by a fast Monte Carlo algorithm developed by Drineas et al.\cite{DKM06}. For a given matrix $\mathbf{A}\in\mathbb{R}^{m\times n}$ and parameters $k_s$, this algorithm returns an approximations to the largest $k_s$ singular values corresponding left singular vectors of the matrix $\mathbf{A}$ in a linear time.

\subsection{Stopping Criterion for Iterations}\label{sec:stop}

We tune the parameters in FPCA for a better overall performance. Continuation is adopted by FPCA, which solves a sequence of instances of \eqref{mtx_cmp}, easy to difficult, corresponding to a sequence of large to small values of $\tau$. The final $\tau$ is the given one but solving the easier instances of \eqref{mtx_cmp} gives intermediate solutions that warm start the more difficult ones so that the entire solution time is reduced. Solving each instance of \eqref{mtx_cmp} requires proper stopping. Because our ultimate goal is to recover 0/1 values on the diagonal of $\mathbf{R}$, accurate solutions of \eqref{mtx_cmp} are not required. Therefore, we use the criterion:
\begin{equation}
\frac{\|\mathbf{M}^{k+1}-\mathbf{M}^{k}\|_{F}}{\max\{1,\|\mathbf{M}^{k}\|_{F}\}}<mtol
\end{equation}
where $mtol$ is a small positive scalar. Experiments shows that $1e^{-6}$ is good enough for obtaining optimal $\mathbf{R}$.

\subsection{Channel Availability Estimation Based on the Complete Measurement Matrix}

Since $\mathbf{F}$ has more columns than rows, directly solving $\mathbf{X}:=\mathbf{R}\mathbf{G}^\top$ in \eqref{eqn:G} from given $\mathbf{M}$ is under-determined. However, each row $X_i$ of $\mathbf{X}$ corresponds to the occupancy status of channel $i$. Ignoring noise in $\mathbf{M}$ for now, $X_i$ contains a positive entry if and only if channel $i$ is used. Hence, most rows of $\mathbf{X}$ are completely zero, so every column $X_{\cdot,j}$ of $\mathbf{X}$ is sparse and all $X_{\cdot,j}$'s are jointly sparse. Such sparsity allows us to reconstruct $\mathbf{X}$ from \eqref{eqn:G} and identify the occupied channels, which are the nonzero rows of $\mathbf{X}$.

Since the channel fading decays fast, the entries of $\mathbf{X}$ have a large dynamic range, which none of the existing algorithms can deal with well enough. Hence, we develop a novel joint-sparsity algorithm briefly described as follows. The algorithm is much faster than matrix completion and typically needs 1-5 iterations. At each iteration, every column $X_{\cdot,j}$ of $\mathbf{X}$ is independently reconstructed using the model $\min\{\sum_{i} w_i |X_{i,j}|: FX_{\cdot,j} = M_{\cdot,j}\}$, where $M_{\cdot,j}$ is the $j$th column of $\mathbf{M}$. For noisy $\mathbf{M}$, we instead use the constraint $\|FX_{\cdot,j} - M_{\cdot,j}\|\le \sigma$. The same set of weights $w_i$ is shared by all $j$ at each iteration. $w_i$ is set to 1 uniformly at iteration 1. After channel $i$ is detected in an iteration, $w_i$ is set to 0. Through $w_i$, joint sparsity information is passed to all $j$. Channel detection is performed on the reconstructed $X_{\cdot,j}$'s at each iteration. It is possible that some reconstructed $X_{\cdot,j}$ is wrong, so we let larger and sparser $X_{\cdot,j}$'s have more say. If there is a relatively large $X_{i,j}$ in a sparse $X_{\cdot,j}$, then $i$ is detected. We have found this algorithm to be very reliable. The detection accuracy is determined by the accuracy of $\mathbf{M}$ provided.

\section{CSS Joint Sparsity Recovery Algorithm}\label{sec:algorithm2}

In this section, we describe a new, highly effective algorithm for recovering
\begin{equation}
\mathbf{X}_{n\times m}=\mathbf{R}_{n \times n}\times (\mathbf{G}_{m \times n})^{\top}
\end{equation}
and thus $\mathbf{R}$ by thresholding $\mathbf{X}$. The algorithm allows but does not require the same $\mathbf{F}$ for all CRs, i.e., each CR can use a different sensing matrix $\mathbf{F}$. The design of $\mathbf{F}$ is discussed in Section \ref{sec:Discussion_design} below.

In $\mathbf{X}$, each column (denoted by $X_{\cdot,j}$) corresponds to the channel occupancy status received by CR $j$, and each row $X_{i,\cdot}$ corresponds to the occupancy status of channel $i$. \emph{Ignoring noise} for now, a row has a positive value (i.e., $|X_{i,\cdot}|>0$) if and only if channel $i$ is used. Since there are only a small number of used channels, $\mathbf{X}$ is sparse in terms of the number of rows containing nonzero. In each nonzero row $X_{i,\cdot}$, there is typically more than one nonzero entry; in other words, if $X_{i,j}\not=0$, other entries in the same row are likely nonzero. Therefore, $\mathbf{X}$ is \emph{jointly sparse}. In the case that the true $\mathbf{X}$ contains noise, it is approximately, rather than exactly, jointly sparse.

Joint sparsity is utilized in our algorithm to recover $\mathbf{X}$. While there are existing algorithms for recovering jointly sparse signals in the literature (e.g., in \cite{DSWBB,TGS05,FR08}), our algorithm is very different and more effective for our underlying problem. None of the existing algorithms works well to recover $\mathbf{X}$ because the entries of $\mathbf{X}$ have a very large dynamic range because, in any channel fading model, channel gains decay rapidly with distance between CRs and PRs. Most existing algorithms are based on minimizing $\sum_i \|X_{i,\cdot}\|_p$ for $p\ge 1$ and $p=\infty$. If $p=1$, it is the same as minimizing the 1-norm of each column independently, so joint sparsity is not
used for recovery. If $p> 1$ or $p=\infty$, joint sparsity is considered, but it penalizes a large dynamic range since the large values in a nonzero row of $\mathbf{X}$ contribute superlinearly, more than the small values in that row, to the minimizing objective. In short, $p$ close 1 loses joint sparsity and $p$ bigger than 1 penalizes large dynamic ranges. Our new algorithm not only utilizes joint sparsity but also takes advantages of the large dynamic range of $\mathbf{X}$.

The large dynamic range has its pros and cons in CS recovery. It makes it easy to recover the locations of large entries, which can be achieved even without recovering the locations of smaller ones. On the other hand, it makes difficult to recover both the locations and values of the smaller entries. This difficulty has been studied in our previous work \cite{MYHH09}, where we proposed a fast and accurate algorithm for recovering 1D signals $x$ by solving several (about 5-10) subproblems in the form of
\begin{equation}\label{trunc}
\mbox{Truncated $\ell_1$ minimization:}\quad\min \{\sum_{i\in T} |x_i|: Ax = b\}
\end{equation}
where the index set $T$ is formed iteratively as $\{1,\ldots,n\}$ excluding the identified locations of large entries of $x$. With techniques such as early detections and warm starts, it achieves both the state--of--the--art speed and least requirement on the number of measurements. We integrate the idea of this algorithm with joint sparsity into the new algorithm below.
\begin{algorithm}
\caption{Joint Detection Algorithm}\label{t:algorithm}
\begin{algorithmic}
\STATE $T \leftarrow \{1,\ldots, n\}$
\REPEAT
\STATE {\textbf{Independence recovery:}}
\STATE $\mathbf{X} \leftarrow 0$
\STATE {$X_{\cdot,j} \gets \min\{\sum_{i\in T} X_{i,j}: A_j X_{\cdot,j} = b_j, X_{\cdot,j}\ge 0 \}$}
%\STATE {
for every CR $j$ with enough measurements (In presence of measurement noise, $A_j X_{\cdot,j} = b_j$ is replaced by $\|A_j X_{\cdot,j} - b_j\|\le \sigma$)
%}
\STATE {\textbf{Channel detection:}}
\STATE {select trusted $X_{\cdot,j}$ and detect used channels from the selections}
\STATE {\textbf{Update of $T$:}}
\STATE {Update $T$ according to detected channels and $\mathbf{X}$}
\UNTIL
{the tail of $\mathbf{X}$ is small enough}
\STATE Report $\mathbf{X}$, and $\mathbf{R}$ by thresholding $\mathbf{X}$
\end{algorithmic}
\end{algorithm}
The framework of the proposed algorithm is shown in Table \ref{t:algorithm}.
At each iteration, every channel is first subject to independent recovery. Unlike minimizing $\sum_i \|X_{i,\cdot}\|_p$, which ties all CRs together, independent recovery allows large entries of $\mathbf{X}$ to be quickly recovered. Joint sparsity information is passed among the CRs through a shared index set $T$, which is updated iteratively to exclude the used channels that are already discovered. Below, we describe each step of the above algorithm in more details.

In the \textbf{independence recovery} step, for every qualified CR, a constrained problem in the form of \eqref{trunc} with constraints $ A_j X_{\cdot,j} = b_j$ in the noiseless case, or $\|A_j X_{\cdot,j} - b_j\|\le \sigma$ in the noisy case, is considered, where $\sigma$ is an estimated noise level. As problem dimensions are small in our application, solvers are easily chosen: MATLAB's `linprog' for noiseless cases and Mosek \cite{MOSEK} for noisy cases. Both of these solvers run in polynomial times. This step dominates the total running time of Algorithm \ref{t:algorithm}, but up to $m$ optimization problems can be solved in parallel.
Parallelization is simple for the joint-sparsity approach. At each outer iteration, all LPs are solved independently, and they have small scales relative to today's LP solvers, like Gurobi \cite{Gurobi} and its MATLAB interface Gurobi Mex \cite{GurobiMex}, where Gurobi automatically detects and uses all CPU and cores for solving LPs.
CRs without enough measurements (e.g., most of their reports are missing due to transmission losses or errors) are not qualified for independent recovery because CS recovery is known unstable in such a case. Specifically, we require the number of the available measurements from each qualified CR to exceed twice as many as used channels or $n-|T|$.

When measurements are ample, the first iteration will yield exact or nearly exact $X_{\cdot,j}$'s. Otherwise, insufficient measurements can cause a completely wrong $X_{\cdot,j}$ that misleads channel detection; neither the locations nor the values of the nonzero entries are correct. The algorithm, therefore, filters trusted $X_{\cdot,j}$'s that must be either sparse or compressible. Large entries in such $X_{\cdot,j}$'s likely indicate correct locations. A theoretical explanation of this argument based on stability analysis for \eqref{trunc} is given in \cite{ISD}.

Used channels are detected among the set of trusted $X_{\cdot,j}$'s. To further reduce the risk of false detections, we compute a percentage for every channel in a way that those channels corresponding to larger values in $\mathbf{X}$ and whose values are located in relatively sparser $X_{\cdot,j}$'s are given higher percentages. Here, relative sparsity is defined proportionally to the number of measurements; for fixed number of non-zeros or degree of compressibility, the more the measurements, the higher the relative sparsity. Hence, $X_{\cdot,j}$ corresponding to more reported CR $j$ also tends to have a higher percentage. In short, larger and sparse solutions have more say. The channels receiving higher percentages are detected as used channels.

The index set $T$ is set as $\{1,\ldots, n\}$ excluding the used channels that are already detected. Obviously, $T$ needs to change from one iteration to the next; otherwise, two iterations will result in an identical $\mathbf{X}$ and thus the stagnation of algorithm. Therefore, if the last iteration posts no change in the set of used channels yet the stopping criterion (see next paragraph) is not met, the channels $i$ corresponding to the larger $\|X_{i,\cdot}\|_2$ are also excluded from $T$, and such exclusion becomes more aggressive as iteration number increases. This is not an ad hoc but a rigorous treatment. It is shown in \cite{ISD} that larger entries in an inexact CS recovery tend to be the true nonzero entries, and furthermore, as long as the new $T$ excludes more true than false nonzero entries by a certain fraction, \eqref{trunc} will yield a better solution in terms of a certain norm. In short, used channels leave $T$, and in case of no leaves, channels with larger joint values $\|X_{i,\cdot}\|_2$ leave $T$.

Finally, the iteration is terminated when the tail of $\mathbf{X}$ is small enough. One way to define the tail size of $\mathbf{X}$ is the fraction $\sum_{i\in T}\|X_{i,\cdot}\|_p / \sum_{i\not\in T}\|X_{i,\cdot}\|_p$, i.e., the thought--unused divided by the thought--used. Suppose that $T$ precisely contains the unused channels and measurements are noiseless, then every recovered $X_{\cdot,j}$ in channel detection is exact, so the fraction is zero; with noise, the fraction depends on noise magnitude and is small as long as noise is small. If $T$ includes any used channel, the numerator will be large whether or not $X_{\cdot,j}$'s are (nearly) exact. In a sense, the tail size measures how well $\mathbf{X}$ and $T$ match the measurements $b$ and expected sparseness. Unless the true number of used channels is known, the tail size appears to be an effective stopping indicator.
\begin{figure}
  \center
  \includegraphics[width=3.7in,height=2.7in]{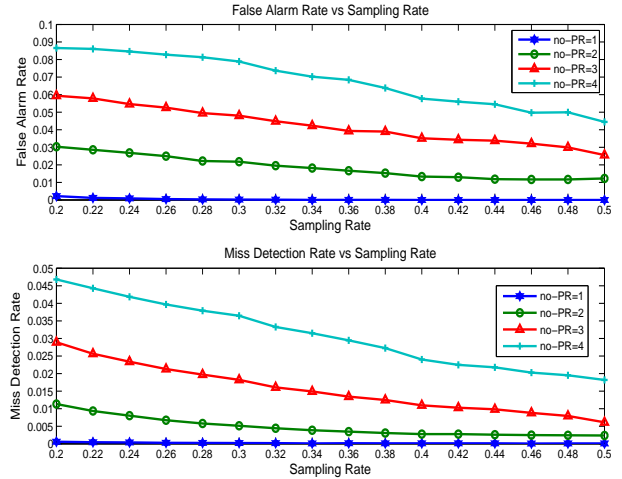}\\
  \caption{False alarm and missing probability vs. sampling rate.}\label{f:FAR_MDR_SR}
\end{figure}
\section{Discussion}\label{sec:Discussion}
\subsection{Complexity}\label{sec:Discussion_complexity}
In the worst case, algorithm \ref{t:algorithm} reduces the cardinality of $T$ by 1 per iteration, corresponding to recovering at least 1 additional used channel. Therefore, the number of iterations cannot exceed the number of total channels. However, the first couple of iterations typically recover most of the used channels. At each iteration, the independence recovery step solves up to $m$ optimization problems, which can be independently solved in parallel, so the complexity equals a linear program (or second-order cone program) whose size is no more than $n$. The worst case complexity is $O(n^3)$ but it is almost never observed in sparse optimization thanks to solution sparsity. The two other steps are based on basic arithmetic and logical operations, and they run in $O(p\times n)$.
In practice, algorithm \ref{t:algorithm} is implemented and run on a workstation at the fusion center. Computational complexity will not be a bottleneck of the system.
As to the matrix completion algorithm, according to \cite{MGC08}, FPCA can recover $1000\times 1000$ matrices of rank 50 with a relative error of $10^{-5}$ in about 3 minutes by sampling only 20 percent of the elements.

\subsection{Comparisons between the Two Approaches}\label{sec:Discussion_compare}
The matrix completion (Section \ref{sec:algorithm1}) and joint sparsity recovery (Section \ref{sec:algorithm2}) approaches both take linear channel measurements as input and both return the estimates of used channels. On the other hand, the joint sparsity approach takes the full advantage of $\mathbf{F}$, so it is expected to work with smaller numbers of measurements. In addition, even though only one matrix completion problem needs to be solved in the matrix completion approach, it still takes much longer than running the entire joint sparsity recovery, and it is not easy to parallelize any of the existing matrix completion algorithms. However, in the small-scale networks, in cases where too much sensing information is lost during transmission or there are too many active PRs in the network, which increase the signal sparsity level, joint sparsity recovery algorithm with our current settings will experience degradation in performance.

We, however, cannot verdict an eventual winner between the two approaches as they are both being studied and improved in the literature. For example, if a much faster matrix completion algorithm is developed which takes advantage of $\mathbf{F}$, the disadvantages of the approach may no longer exist.

\subsection{Frequency-Selective Filter Design and Adaptive Sensing}\label{sec:Discussion_design}
The proposed method senses the channels, not by measuring the responses of individual channels one by one, but rather measures a few incoherent linear combinations of all channels' responses through onboard frequency-selective filter set. The filter coefficients which perform as the sensing matrix should have entries independently sampled from a sub-gaussian distribution, since this is known to be best for compressive sensing in terms of the number of measurements (given in order of magnitude) required for exact recovery. In other words, up to a constant, which is independent of problem dimensions, no other type of matrix is yet known to perform consistently better. However, other types of matrices (such as partial Fourier/DCT matrices \cite{CT06, CRT06} and other random circulant matrices \cite{YMYZ10}) have been either theoretically and/or numerically demonstrated to work as effectively in many cases. These latter sensing matrices are often easier to realize physically or electrically. For example, applying a random circulant matrix performs sub-sampled convolution with a random vector.

Frequency-selective surfaces (FSSs) can be used to realize frequency filtering. This can be done by designing a planar periodic structure with a unit element size around half wavelength of the frequency of interests. Both the metallic and dielectric materials can be used. To deal with the bandwidth, unit elements in different shapes will be tested.

\subsection{Dynamic CS Update}\label{sec:Dynamic_CS_Update}
Channel occupancy evolves over time as PRs start and stop using their channels. Channel gains can also change when the PRs move. However, the CS research has so far focused on static signal sensing except the very recent path following algorithms in \cite{AJ091,AJ092}. In the future work, we can investigate recovery methods for a dynamic wireless environment where based on existing channel occupancy information, an insignificant change of channel states can be quickly and reliably discovered.
Given existing channel occupancy $\mathbf{X}$, each new report, which is an entry ${M}_{i;j}$ of $\mathbf{M}$, is compared with $\mathbf{(FX)}_{i;j}$.
If a significant number of such comparisons show differences, then there is a change in the true $\mathbf{X}$. Since $\mathbf{X =(RG^T)}$, either $\mathbf{R}$ or $\mathbf{G}$, or both, have changed. A change in $\mathbf{R}$ means new channel occupation or release. If $\mathbf{R}$ is unchanged, then those channel gains in $\mathbf{G}$ corresponding to occupied channels have changed.
It is easy to deal with the latter case (i.e., $\mathbf{G}$ changed, but $\mathbf{R}$ didn't) and update the gains of occupied channels because it boils down to solving a small linear system. Let $\mathbf{\hat{F}}$ and $\mathbf{\hat{X}}$ denote the sub-matrices of $\mathbf{F}$ and $\mathbf{X}$, respectively, formed by their columns and rows corresponding to the occupied channels. Then, the new gains are given in the least-squares solution of $\mathbf{M} = \mathbf{\hat{F} \hat{X}}$, where $\mathbf{M}$ shall include new reports arrived after the previous recovery/update but may still have missing entries.
This system is easy to solve since the number of occupied channels is small.

In a similar way it is easy to discover released channels as long as there is no introduction of new occupied channels. The release of channel $i$ means row $X_i$ of $\mathbf{X}$ turns into 0, or small numbers. Therefore, one can solve the system $\mathbf{M} = \mathbf{\hat{F} \hat{X}}$ and find the released channels, which correspond to the rows of $\mathbf{\hat{X}}$ with all zero (or small) entries.
When the system $\mathbf{M} = \mathbf{\hat{F} \hat{X}}$ is inconsistent, it means that the received reports cannot be explained by the previously occupied channels, so there must be new channel occupation. Discovering new channel occupation is more difficult since it is to find changes in the previously unoccupied ones, which
are much more than the occupied channels. However, it is computationally much easier than starting from scratch. Let $\mathbf{X}_{prev}$ and $\mathbf{X}$ denote the previous and current channel information, respectively.
Arguably, $\mathbf{X}_{prev}-\mathbf{X}$ is highly sparse in the joint sense because only its rows corresponding to newly occupied or released channels can have large nonzero entries. Hence, $\mathbf{X}$ can be quickly recovered by performing joint sparsity recovery on $\mathbf{X}_{prev}-\mathbf{X}$ over the constraints $\mathbf{M} = \mathbf{FX}$ (or a relax version in the noisy case), a task that can be done by the algorithms for stationary recovery.

\section{Simulation Results}\label{sec:simulation}
The Probability of Detection (POD) and False Alarm Rate (FAR) are the two most important indices  associated with spectrum sensing. We also consider the Miss Detection Rate (MDR) of the proposed system. The higher the POD, the less interference will the CRs bring to the PRs, while from the CRs' perspective, lower FAR will increase their chance of transmission. There is a tradeoff between POD and FAR. While designing the algorithms, we try to balance the CR nodes' capability of transmission and their interferences to the PR nodes. Performance is evaluated in terms of POD, FAR and MDR defined as follows:
\begin{equation}\nonumber
\begin{array}{lll}
\mbox{FAR=No. \ False /(No.\ False+No.\ Hit)}\\
\mbox{MDR=No.\ Miss/(No.\ Miss+No.\ Correct)}\\
\mbox{POD=No.\ Hit/(No.\ Hit+No.\ Miss)}\\
\end{array}
\end{equation}
where \emph{No. False} is the number of false alarms, \emph{No. Miss} is the number of miss detections, \emph{No. Hit} is the number of successful detections of primary users, and \emph{No. Correct} is the number of correct reports of no appearance of PR.
We define sampling rate as
\begin{equation}\nonumber
%\begin{array}{lc}
\frac{\mbox{No.}\ \mbox{received}\ \mbox{measurements}\  \mbox{at}\  \mbox{the}\  \mbox{fusion}\  \mbox{center}}{\mbox{No.}\ \mbox{channels} \times \mbox{No.}\ \mbox{CRs}}
%\end{array}
\end{equation}
where $\mbox{(No.\ channel} \times \mbox{No.\ CR})$ is the amount of total sensing workload in traditional spectrum sensing.
\begin{figure}
  \center
  \includegraphics[width=3.7in,height=2.7in]{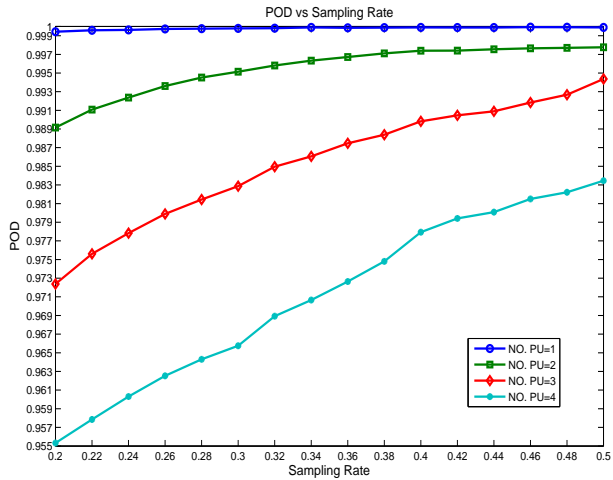}\\
  \caption{POD vs. sampling rate.}\label{f:POD_SR}
\end{figure}

\subsection{Simulation of Matrix Completion Recovery}\label{sec:simulation_MC}

According to FCC and Defense Advance Research Projects Agency (DARPA) reports \cite{NTIA,FCC} data, we chose to test the proposed matrix completion recovery algorithm for spectrum utilization efficiency over a range from 3\% to 12\%, which is large enough in practice. Specifically, the number of active primary users is 1 to 4 on a given set of 35 channels with 20 CR nodes.

Fig. \ref{f:FAR_MDR_SR} shows the false alarm and miss detection rates at different sampling rates for different numbers of PR nodes. Among all cases, the highest miss detection rate is no more than 5\%, and this is from only 20\% samples which are supposed to be gathered from the CR nodes regarding all the channels. When the sampling rate is increased to 50\% and even when the channel occupancy is relatively high, i.e., 12\% of the channels are occupied by the PRs, the miss detection rates can be as low as no more than 2\%. From our simulation results, with a moderate channel occupancy at 9\%, the false alarm rates are around 3\% to 5\%.
\begin{figure}
\center
\includegraphics[width=3.7in,height=2.7in]{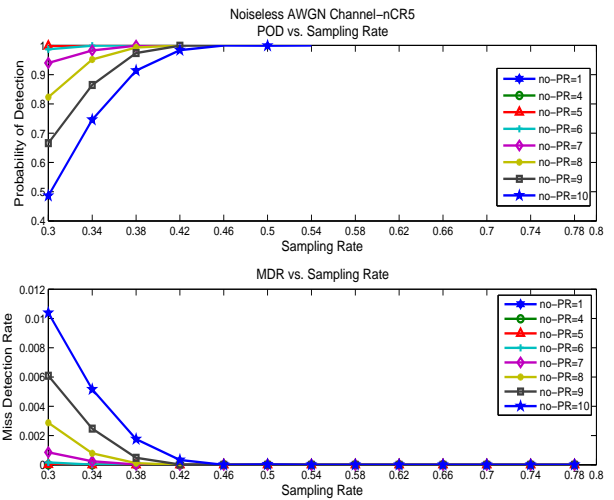}\\
\caption{Noiseless AWGN channel (no. of CR = 5).}\label{f:NN_T0_5}
\end{figure}
Fig. \ref{f:POD_SR} shows the probability of detection at different sampling rates. When the spectrum is lightly occupied by the licensed user at 3\% channels being occupied, only 20\% samples offer a POD close to 100\%, and when there is a slightly raise in sampling rate, POD can reach 100\%. In the worst case of 12\% spectrum occupancy, 20\% sampling rate still can offer a POD of higher than 95\%, and as the sampling rate reaches 50\%, POD can reach 98\%.

\subsection{Joint Sparsity Recovery Simulation}\label{sec:simulation_JS}

\begin{figure}
\center
\includegraphics[width=3.7in,height=2.7in]{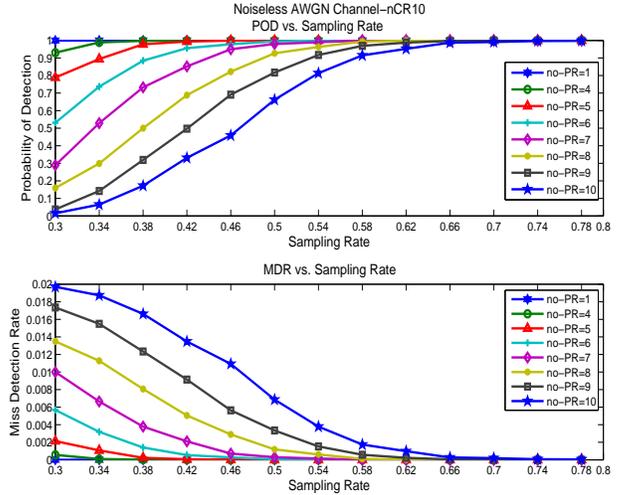}\\
\caption{Noiseless AWGN channel (no. of CR = 10).}\label{f:NN_T0_10}
\end{figure}

\begin{figure}
\center
\includegraphics[width=3.7in,height=2.7in]{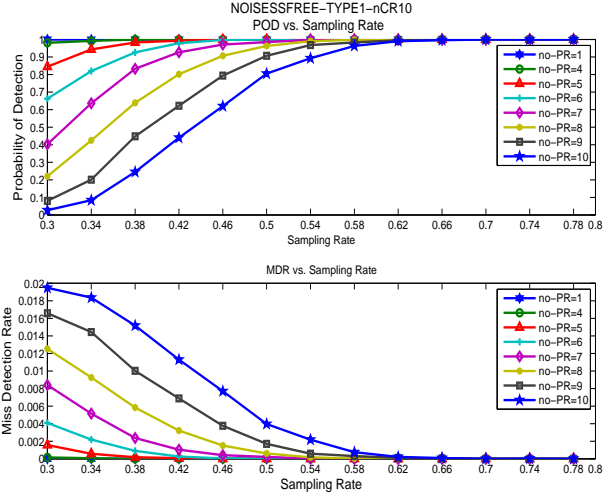}\\
\caption{Noiseless Rayleigh fading channel (no. of CR = 10).}\label{f:NN_T1_10}
\end{figure}

\begin{figure}
\center
\includegraphics[width=3.7in,height=2.7in]{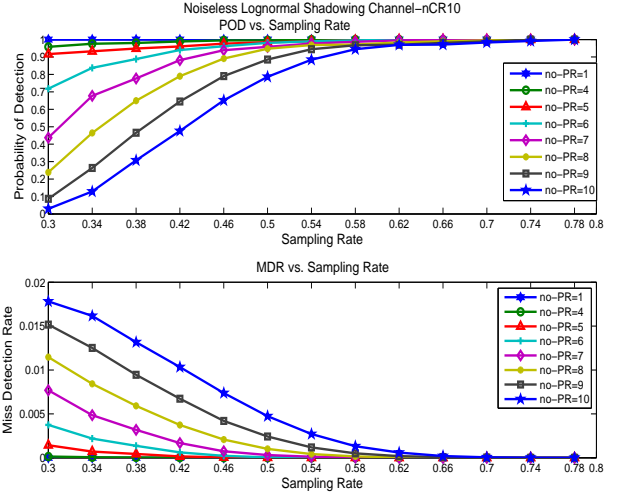}\\
\caption{Noiseless log-normal shadowing channel (no. of CR = 10).}\label{f:NN_T2_10}
\end{figure}

\begin{figure}
\center
\includegraphics[width=3.7in,height=2.7in]{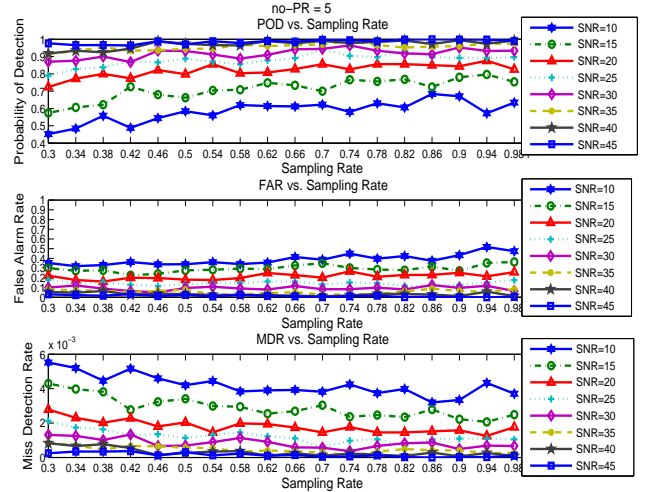}\\
\caption{POD, FAR, and MDR performance vs. sampling rate at different SNR.}\label{f:N_Performance}
\end{figure}

\begin{figure}
\center
\includegraphics[width=3.7in,height=2.7in]{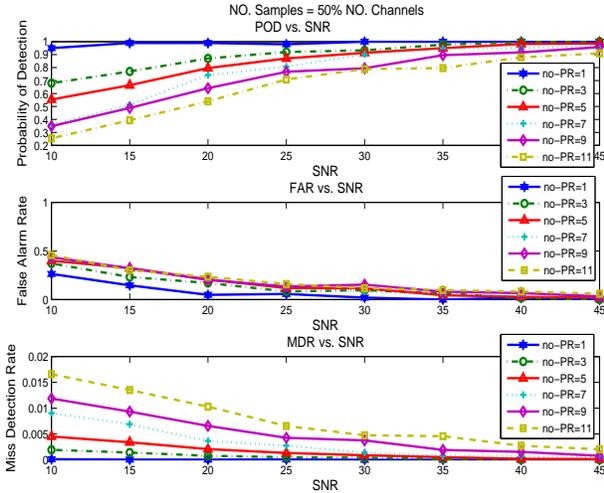}\\
\caption{POD, FAR, and MDR performance vs. noise level for different number of PR.}\label{f:N_Per_SNR}
\end{figure}

Joint sparsity recovery is designed for large scale application, and simulations carried out for a larger dimensional applications with the following settings:
We consider a $20$-node cognitive radio network within a $500\times 500$ meter square area centered at the fusion center. The $20$ CR nodes are uniformly randomly located. These cognitive radio nodes collaboratively sense the existence of primary users within a $1000\times 1000$ meter square area on $500$ channels, which are centered also at the fusion center. We chose to test the proposed algorithm for the number of active PR nodes ranging from $1$ to $15$ on the given set of 500 channels.
Since the fading environments of the cognitive radio networks vary, we evaluate the algorithm performance under three typical channel fading models: AWGN channel, Rayleigh fading channel, and lognormal shadowing channel.
\begin{figure}
\center
\includegraphics[width=3.7in,height=2.7in]{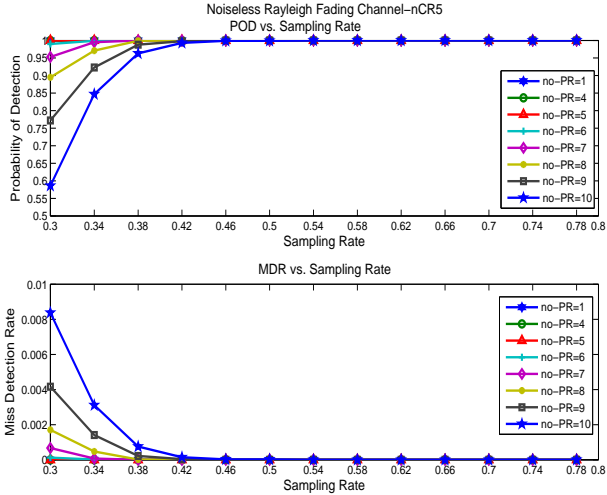}\\
\caption{Noiseless Rayleigh fading channel (no. of CR = 5).}\label{f:NN_T1_5}
\end{figure}
\begin{figure}
\center
\includegraphics[width=3.7in,height=2.7in]{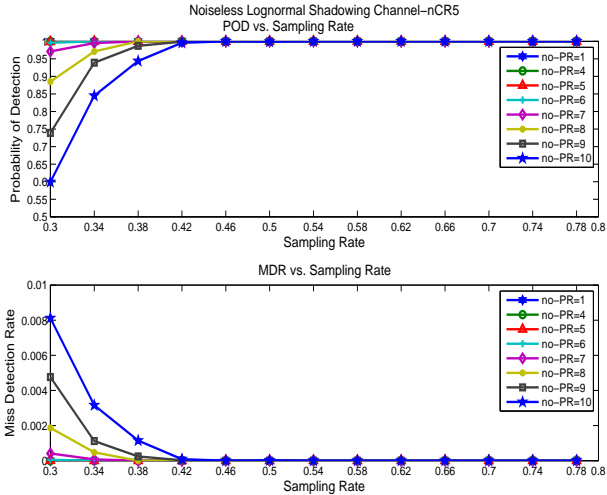}\\
\caption{Noiseless log-normal shadowing channel  (no. of CR = 5).}\label{f:NN_T2_5}
\end{figure}
We first evaluate the POD, FAR, and MDR performance of the proposed joint sparsity recovery performance in the noiseless environment. Fig. \ref{f:NN_T0_5}, Fig. \ref{f:NN_T1_5}, and Fig. \ref{f:NN_T2_5} show the POD, FAR and MDR performance at different sampling rate, for AWGN channel, Rayleigh fading channel, and lognormal shadowing channel, respectively, when small number of CR nodes sense the spectrum collaboratively. Fig. \ref{f:NN_T0_10}, Fig. \ref{f:NN_T1_10}, and Fig. \ref{f:NN_T2_10} show the POD, FAR and MDR performance at different sampling rate, for the aforementioned three types of channel models, when there are more CR nodes involved in the collaborative sensing of the spectrum. We observe that, log-normal shadowing channel model shows the best POD, FAR, and MDR performance no matter how many CR nodes are involved in the spectrum sensing. While the AGWN channel model shows the worst POD, FAR, and MDR performance. With respect to POD, the performance gap between these two models is at most 10\%, which happens when the sampling rate is extremely low. For the Rayleigh fading channel model, when the number of samples is $62\%$ of the total number of channels, for all tested cases we achieve $100\%$ POD. If there are less active PR nodes in the network, smaller number of samples are required for exact detection. In essence, the proposed CCS system is robust to severe or poorly modeled fading environments. Cooperation among the CR nodes and robust recovery algorithm allow us to achieve this robustness without imposing stringent requirements on individual radios.
We then evaluate the POD, FAR, and MDR performance of the proposed joint sparsity recovery performance in noisy environments. For all the  simulations considering noise, we adopt the Rayleigh fading channel model. Fig. \ref{f:N_Performance} and Fig. \ref{f:N_Per_SNR} show the corresponding results. We observe that noise does degrade the performance. However, as shown in Fig. \ref{f:N_Performance}, when the number of active PRs is small enough (e.g., no. of PR = 1), even with signal to noise ratio as low as 15 dB, we still can achieve $100\%$ POD with a sampling rate of merely $50\%$. Then with an increase in the signal to noise ratio, lower sampling rate enables more PR nodes to be detected exactly. Fig. \ref{f:N_Per_SNR} shows the POD, FAR and MDR performance vs. sampling rate at different noise level, each curve for a specific noise level is relatively flat (i.e., performance varies a little as sampling rate changes). This shows that the noise level has greater impact on the spectrum sensing performance rather than the sampling rate. At low noise level, e.g., SNR = 45 dB, $40\%$ sampling rate enables $100\%$ POD for 4 PR nodes. As SNR reduces to 15 dB, no more than $70\%$ POD will be achieved even when the number of samples equals to the number of channels in the network.
\subsection{Comparison between Matrix Completion Algorithm and Joint Sparsity Recovery Algorithm}
For comparison, we applied joint sparsity recovery algorithm on a small-scale network with the same settings as we have used to test the matrix completion recovery. Instead of using a 500-channel network, we use a network with only 35 channels. Simulation results show that joint sparsity recovery algorithm performs better than the matrix completion algorithm in the following aspects:
 \begin{enumerate}
   \item Faster computation due to lower computational complexity;
   \item Higher POD for the spectrum utilization rate between 3\% and 12\% in the noise free simulations;
 \end{enumerate}

 To conclude, matrix completion algorithm is good for small-scale networks, with relatively high spectrum utilization, while joint sparsity recovery algorithm has the advantage of low computational complexity which enables fast computation in large-scale networks.
\section{Conclusions}\label{sec:conclusion}

In order to reduce the amount of sensing and transmission overhead of cognitive radio (CR) nodes, we have applied compressive sensing for collaborative spectrum detection in cognitive radio networks. We propose to equip each CR node with a frequency-selective filter, which linearly combines multiple channel information, and let it send a small number of such linear combinations to the fusion center, where the channel occupancy information is then decoded. Consequently, the amount of channel sensing at the CRs and the number of reports sent from the CRs to the fusion center reduce significantly.

Two novel decoding approaches have been proposed -- one based on matrix completion and the other based on joint sparsity recovery. The novel matrix completion approach recovers the complete CR--to--center reports from a small number of valid reports and then reconstructs the channel occupancy information. The joint sparsity approach, on the other hand, skips recovering the reports and directly reconstructs channel occupancy information by exploiting the fact that each occupied channel is observable by multiple CR nodes. Our algorithm enables faster recovery for large-scale cognitive radio networks.

The primary user detection performance of the proposed approaches has been evaluated by simulations. The results of random tests show that, in noiseless cases, the number of samples required are no more than 50\% of the number of channels in the network to guarantee exact primary user detection for both approaches; while in noisy environments, at low channel occupancy rate, we can still have high probability of detection.

\section*{Acknowledgements}
The work of W. Yin was supported in part by NSF CAREER Award DMS-07-48839, ONR Grant N00014-08-1-1101, and an Alfred P. Sloan Research Fellowship.
The work of H. Li was supported in part by NSF Grants 0831451 and 0901425. The work of Z. Han was supported in part by NSF CNS-0910461, CNS-0901425, NSF CAREER Award CNS-0953377, and Air Force Office of Scientific Research. The work of E. Hossain was supported by the NSERC, Canada,  strategic project grant STPGP 380888.

\bibliographystyle{IEEEbib}

\end{document}